\documentclass[twocolumn,prl,preprintnumbers,amsmath,amssymb,letterpaper]{revtex4}

\newcommand{\be}
{\begin{eqnarray}}

\newcommand{\ee}
{\end{eqnarray}}

\usepackage{graphicx}
\usepackage{dcolumn}
\usepackage{bm}
\usepackage{times}

\begin{document}

\bibliographystyle{unsrt}

\title{Non-Markovian decoherence of localized nanotube excitons by acoustic phonons}
\author{Christophe Galland}
\author{Alexander H\"{o}gele}
\author{Hakan E. T\"{u}reci}
\author{Ata\c{c} Imamo\u{g}lu}
\affiliation{Institute of Quantum Electronics, ETH Z"{u}rich, Wolfgang-Pauli-Strasse 16, CH-8093
Z\"{u}rich, Switzerland}

\date{\today}

\begin{abstract}
We demonstrate that electron-phonon interaction in quantum dots embedded in one-dimensional systems
leads to pronounced, non-Markovian decoherence of optical transitions. The experiments we present
focus on the line shape of photoluminescence from low-temperature axially localized carbon nanotube
excitons. The independent boson model that we use to model the phonon interactions reproduces with
very high accuracy the broad and asymmetric emission lines and the weak red-detuned radial
breathing mode replicas observed in the experiments. The intrinsic phonon-induced pure dephasing of
the zero-phonon line is two orders of magnitude larger than the lifetime broadening and is a
hallmark of the reduced dimensionality of the phonon bath. The non-Markovian nature of this
decoherence mechanism may have adverse consequences for applications of one-dimensional systems in
quantum information processing.
\end{abstract}

\maketitle

In single-wall carbon nanotubes (SWNTs), the one-dimensional (1D) nature of electronic states
results in an enhancement of Coulomb interactions and has lead to the observation of remarkable
effects such as Luttinger-liquid states in transport experiments \cite{TLLlowT}. On the other hand,
a common feature of the small diameter semiconducting SWNTs is the unintentional confinement of the
electronic or excitonic states along the nanotube axis; this leads to the formation of
zero-dimensional (0D) quantum dot (QD)-like states, manifesting themselves through Coulomb blockade
in transport experiments \cite{coulomb_blockade,coulomb_blockadeSC} or through localized emission
\cite{nanoscaled_PL} and strong anti-bunching in photoluminescence (PL) \cite{hogele}. Spatial
confinement of electrons or excitons in SWNTs is in fact desirable for applications in single
photonics or quantum information processing (QIP), which in turn motivates theoretical and
experimental investigations of elementary physical properties of optically active QDs in SWNTs.

Electron coupling to Raman-active phonon modes such as radial breathing mode (RBM) and G mode has
been an invaluable tool for characterizing SWNTs. Interest in long-wavelength acoustic phonons of
SWNTs primarily stems from their potential use as nano-mechanical resonators. A question that is of
interest from QIP perspective is the decoherence induced by interactions between QD electrons or
excitons and 1D acoustic phonons: while this problem has been addressed theoretically for QD
excitons in quasi-1D III-V nanowires in the high temperature limit \cite{qwire} experimental
confirmation of predicted features have not been obtained.

In this letter, we analyze the influence of exciton-phonon interactions on the PL line shape of
SWNT QDs. We show that the experimental PL spectra consisting of broad and strongly asymmetric
lines can be fit with remarkable accuracy using an analytical solution to the independent boson
model, over a dynamical range spanning two orders of magnitude and a temperature range varying from
5K to 35K. In our model, we assume that the lowest bright SWNT excitons, formed out of bound
electron-hole pairs from the first band-to-band transition $E_{11}$ \cite{exc_chiral}, are confined
to form QD-like states; this assumption is justified since PL has been shown to exhibit strong
photon antibunching \cite{hogele}. The principal result of our work is the demonstration that 1D
nature of acoustic phonons and the resulting Ohmic exciton-phonon coupling gives rise to strong
pure dephasing of spatially localized optical excitations even as we approach the zero-temperature
limit. Even though we study localized SWNT excitons in our experiments, we emphasize that our
conclusions are valid for localized transitions in any 1D system. Our findings also suggest that
simple PL measurements can be used to determine quantities such as the localization length scale of
excitons and the deformation potential coupling strength in carbon nanotubes.

We measured the PL  from isolated SWNTs at cryogenic temperatures. Details on the experimental
setup and sample preparation are given in \cite{hogele}. Briefly, we deposed surfactant-embedded
CoMoCat SWNTs directly on a functionalized solid-immersion lens (SIL) and used a confocal
microscope to detect PL upon excitation with a tunable Ti:Sapphire  laser (CW or pulsed mode). The
sample is placed in a helium bath cryostat and the temperature is controlled locally through a
thermo-resistor. As discussed in \cite{hogele}, spatially localized emission and strong photon
anti-bunching indicate that the PL signal originates from unintentionally axially confined
excitons, i.e. from SWNT QDs. Figure~1 shows typical experimental PL spectra (open circles) at $T =
4.2$~K that we observed: the line shape is clearly asymmetric with a full-width-half-maximum (FWHM)
 of 3.5~meV. Time-resolved measurements indicate that
the PL decay follows a bi-exponential law with a fastest time constant of 36 psec (Fig.~1 inset) in
agreement with previous reports (see \cite{Hagen,Hirori} and references therein), which implies
that the measured linewidth is a factor of 100 larger than the lifetime broadening. This particular
nanotube PL exhibited strong photon antibunching with a value for the normalized second-order
correlation function $g^{(2)}(0) = 0.1$ \cite{hogele}.

\begin{figure}[t]
\includegraphics[scale=0.8]{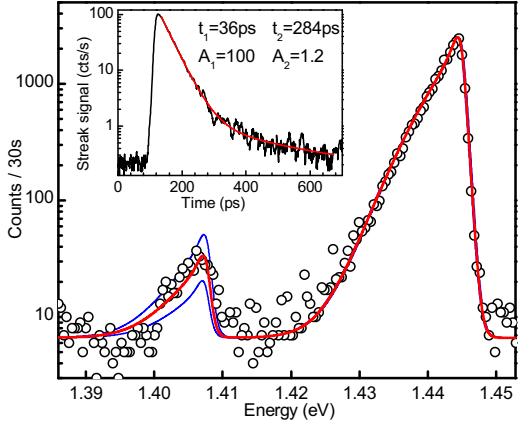}
\caption{Photoluminescence of a nanotube quantum dot at T = 4.2 K and P = 300 nW (open circles) and
the calculated spectrum (solid line) plotted on a logarithmic scale. For the theoretical model we
used the parameters $D_{S}=14$~eV, $D_{RBM} = 1.4$~eV/{\AA}, $\hbar \omega_{RBM} = 37.2$~meV,
$\Delta E = 44$~meV  and $T_{fit}= 6.3$~K.
The upper and lower blue solid lines are calculated spectra with
identical parameters but $ D_{RBM} = 1.8$~eV/{\AA} and $ D_{RBM} = 1.0$~eV/{\AA}, respectively,
demonstrating the precision with which the coupling strength can be determined. The inset shows a
time-resolved measurement (dark line)
The red line is a bi-exponential fit of the PL decay revealing a main fast component with
characteristic time $t_1= 36 \pm 0.2$~ ps and weight $A_1=100 \pm 0.3 $ and a vanishingly weak long
tail ($t_2 = 284 \pm$~ 40 ps , $A_2 = 1.2 \pm 0.2 $). From the value of $t_1$ we expect a
lifetime-limited line-width of $\sim 0.036$~ meV, in strong contrast to the measured and calculated
width of 3.5~meV. } \label{fig1}
\end{figure}

To explain this broad and asymmetric PL line shape we develop a theoretical model where we assume a
harmonic confinement potential for excitons along the nanotube axis. For confinement length scales
that are long compared to the exciton Bohr radius, QD states are superpositions of free excitonic
states with envelope wave functions determined by the 1D harmonic oscillator Hamiltonian. We
further assume that the confinement is strong enough to lead to well-separated QD levels, with an
energy splitting $\Delta E \gg k_B T$, where $k_B$ is the Boltzmann constant and $T$ the absolute
temperature.

To account for the coupling of this QD to the acoustic phonons of the SWNT, we follow the extension
of the independent boson model developed in \cite{qdtheo}. We assume that the phonon dispersion of
the bare SWNT remains unaltered by the confining potential. Since no atomic bonds are formed with
the embedding surfactant, we also assume that coupling to phonons of the surrounding material is
negligible. Long wavelength acoustic phonons in nanotubes are well described by a continuum elastic
model \cite{CNT_phonon_cont, CNT_phonon_transp} and they couple to electrons through deformation
potentials \cite{CNT_phonon_def,CNT_phonon_cont, CNT_phonon_transp}. Because of momentum
conservation, only phonon modes with zero angular momentum can couple to excitonic states within a
single subband $E_{11}$; the potentially relevant modes are therefore stretching, twisting and
radial breathing modes. We neglect the coupling to the twisting mode, whose strength is expected to
be one order of magnitude smaller than to the stretching mode \cite{CNT_phonon_cont}. The
stretching mode corresponds to lattice vibrations with longitudinal displacements of the atoms
along the nanotube axis \cite{CNT_phonon_transp}. It has linear dispersion $\omega_s(q)=v_s q$ for
small wave-vectors \cite{CNT_phonon_transp} and the strength of its coupling to excitons is
characterized by the deformation potential $D_{S}$. It would provide the dominant contribution to
the broadening of the localized exciton emission, a.k.a. the zero-phonon line (ZPL). Since the RBM
has an approximately flat dispersion for small wave-vectors (with an energy $\hbar \omega_{RBM}$),
its principal contribution would be the appearance of phonon sidebands in PL that are separated
from the exciton line by $\hbar \omega_{RBM}$.

Following ref \cite{qdtheo}, we obtain the exciton-phonon matrix elements:
$g_j(q)=\mathcal{G}_j(q)\cdot \mathcal{F}^{exc}(q)$
where $j$ stands for $S$ (stretching mode) or $RBM$. The form factor $\mathcal{F}^{exc}(q)=\int dz
|\Psi^{exc}(z)|^2 e^{i q \cdot z} $ provides a cut-off for the coupling in momentum space that is
inversely proportional to the spatial extension of the wave function.
With our choice of a parabolic potential, we get for the QD ground state a gaussian envelope along
$z$: $\Psi^{exc}(z)=\frac{1}{\pi^{1/4} \sigma^{1/2}} \exp{-\frac{z^2}{2\sigma^2}} $ where $\sigma$
is the confinement length. This yields $\mathcal{F}^{exc}(q)=\exp{-\frac{q^2 \sigma^2}{4}}$ and the
energy splitting: $\Delta E = \frac{\hbar^2}{m^*_{exc} \sigma^2}$, where $m^*_{exc}$ is the
effective mass of the exciton. The deformation potential couplings are:
  $\mathcal{G}_S(q) = \frac{D_s \cdot q}{\sqrt{2 \rho L \hbar \omega_s(q)}} $ and
  $\mathcal{G}_{RBM}(q) = \frac{D_{RBM}}{\sqrt{2 \rho L \hbar \omega_{RBM}}}$
where $L$ is the length of the nanotube and $\rho$ its linear mass density.

The linear susceptibility $\chi(t)$ ($t \geq 0$) of the QD in response to a $\delta$-shaped laser
pulse at $t=0$  \cite{qdpulse, qdtheo} can be decomposed into a temperature-dependent  and a
temperature-independent contribution $\chi(t) = -i e^{-i \overline{\Omega} t} \chi_{T}(t) \cdot
\chi_{0}(t)$ with
\vspace{-5pt}
\begin{eqnarray}
    \chi_{T}(t)  & \propto & i  \exp  \sum_{q} |\gamma_S(q)|^2 [ - n_S(q)  |e^{-i \omega_S(q) t}-1|^2  ]  \label{eqchiT}\\
    \chi_{0}(t) & \propto & i  \exp  \sum_{q} |\gamma_S(q)|^2 [ e^{-i \omega_S(q)t}-1] \label{eqchiT0}
\end{eqnarray}
\vspace{-10pt}

\noindent where $\gamma_j(q)=g_j(q) / \omega_j(q)$, and $n_j(q)= \left( e^{\hbar \omega_j(q) / k_B
T} -1 \right)^{-1} $ is the phonon occupation number. The polaron-shifted transition energy is
$\overline{\Omega}=\Omega-\sum_{j,q} |\gamma_j(q)|^2 \omega_j(q)$ , where $\hbar \Omega$ is the
bare energy of the QD-state. Performing the Fourier transform of $\chi(t)$ and taking  its
imaginary part gives the absorption spectrum of the QD. The PL-emission line shape that we measure
experimentally is obtained by taking the mirror image of the absorption profile with respect to the
ZPL \cite{mahan}.

\begin{figure}[t]
\includegraphics[scale=0.75]{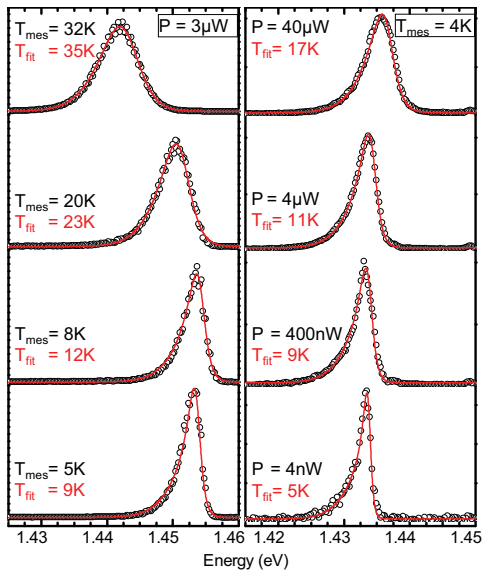}
\caption{Left panel: Experimental (open circles) and calculated (red solid lines) PL spectra for
the same SWNT at different temperatures $T_{mes}$ (the vertical scale is arbitrary). The excitation
power was kept constant at $P = 3\mu W $. The effective temperature used in the simulation is given
by $T_{fit}$, the other parameters are $ D_{S} = 12.7$~ eV and $ \Delta E = 22$~ meV. Right panel:
Emission from another nanotube at constant sample temperature T = 4.2~K. The changes induced by
increasing the laser intensity from $4$~nW to $40$~$\mu$W are very well reproduced by increasing
the temperature $T_{fit}$ in the model. The parameters for this SWNT are $D_{S}=13$~ eV and $\Delta
E = 20$~ meV . } \label{fig2}
\end{figure}

\begin{figure}[t]
\includegraphics[scale=0.7]{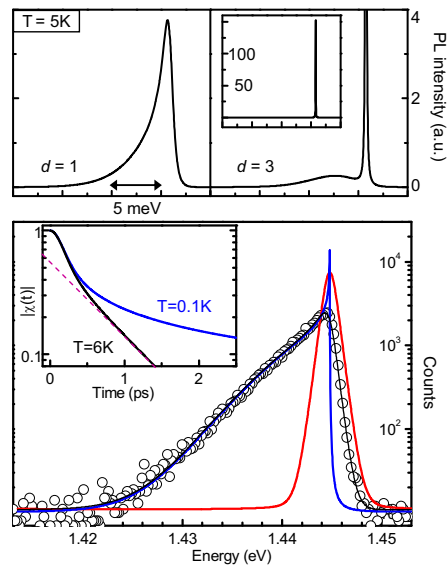}
\caption{Upper panel: Calculated emission line-shapes at $T=5$ K for a QD coupled to a phonon bath
of dimension $d = 1$ (left) and $d=3$ (right) are shown. The coupling to a single acoustic phonon
branch with linear dispersion ($v_s = 19.9$~Km/s) is considered.  To extend the model to higher
dimension, we assume an isotropic coupling of the QD to LA phonons described by the same
deformation potential. A lifetime broadening corresponding to $t_1 = 36$~ps is introduced. Lower
panel: For illustration, fit of the spectrum shown in Fig.~1 (open circles) with only the
temperature dependent (Eq.~\ref{eqchiT}, red line) and the temperature-independent
(Eq.~\ref{eqchiT0}, blue line) contributions. Each curve is scaled to span the same total area. The
inset shows the susceptibility ($T = 6$ K) used to fit the data (black curve), where a deviation
from an exponential decay is visible for $t \le 1$psec. The blue curve of the inset shows the same
function at $T=0.1$ K where the non-Markovian nature of dephasing becomes clearer.}
\label{fignonmarkov}
\end{figure}

The solid curve in Fig.~1 shows that this simple theoretical model describes the experimental
observation remarkably well. To obtain this fit, we use SWNT parameters that were reported earlier;
in particular, we take $\rho = 1.67 \cdot 10^{-15}$~kg/m, $v_s=19.9$~Km/s \cite{CNT_phonon_cont}
and $m^*_{exc}\simeq 0.2 \cdot m_e$ \cite{exc_effects}. An electron-phonon deformation potential
for the stretching mode \cite{CNT_phonon_cont,CNT_phonon_transp} of strength $D_S =14$~eV has been
extracted from data on low-field mobilities in semiconducting SWNTs \cite{CNT_phonon_def}
\footnote[1]{We note that this is a measure of the electron-phonon coupling, whereas the relevant
value in our case is the difference between the electron and hole deformation potentials, which has
never been measured so far.}. Our procedure, applied to 10 different SWNT spectra, involves varying
the values of $D_S$ and the confinement to obtain the best fit to the data. The range of $D_S$
values we thereby obtained varied from 12 to 14~eV. The other fitting parameter is the confinement
potential along the SWNT axis which determines the cut-off in momentum space, or equivalently the
level spacing $\Delta E$. We found values for $\Delta E$ between 20 and 44~meV, consistent with our
assumption $\Delta E \gg k_B T \approx 2.8$~meV (at 35~K) and corresponding to an extension of the
wave function envelope on the order of $\sim 10$~nm. Since this confinement lengthscale is bigger
than the Bohr radius \cite{exc_chiral,Maultzsch}, our assumption of exciton center-of-mass
confinement is also justified.

The PL  spectrum from Fig.1 is plotted on a logarithmic vertical scale to show the quality of the
fit over two decades of counts and to emphasize the presence of a red-detuned replica that is
exactly reproduced by the model. The RBM deformation potential for excitons are well documented
\cite{eph_matrix_elt,eph_raman1,eph_raman2} and expected to lie between 1 and 2~eV/{\AA}. Here we
obtained the best fit for $D_{RBM}=1.4 \pm 0.2$~eV/{\AA} and a RBM energy of 37.2~meV corresponding
to a SWNT diameter of about 0.7~nm, in good agreement with CoMoCat SWNTs emitting at this
wavelength. The linear power dependence of the red-detuned replica for very low excitation
intensities (not shown) is consistent with that of a phonon sideband and excludes an explanation
based on biexciton emission.

Figure~2 shows the experimental PL lines (open circles) for two different SWNTs. In the left panel,
the temperature is increased by flowing current in the thermo-resistor. The very asymmetric lines
at the lowest temperatures and the evolution to broader symmetric lines with increasing temperature
is a general feature common to virtually all of the 30+ SWNTs that we measured. The computed
spectra (solid red lines) reproduce both qualitatively and quantitatively this behavior when the
coupling to the stretching mode only is taken into account. Only the considerable shifts in the
transition energies are not reproduced by the model; arbitrary offsets were therefore introduced
for each spectrum. This unexpected sensitivity on minor environmental changes is another evidence
for localized emitting states \cite{hogele}. In contrast, the transition energy of a delocalized
exciton is not expected to shift so dramatically  with temperature \cite{exc_shift}.

The right panel in Fig.~2 (different SWNT) shows that the effect of increasing the excitation power
can be reproduced by only increasing the temperature of the phonon bath in the model and keeping
all the other parameters fixed. From this, one can conclude that the laser excitation led a to a
very local heating of the SWNT that was not directly measured by our thermo-resistor placed about
$1$~mm away from the nanotubes.

We emphasize that the unusually broad lines observed in SWNT PL can be accurately and
quantitatively explained by solely taking into account interaction with long wavelength 1D acoustic
phonons. The resulting ohmic coupling \cite{Wilson} gives rise to linewidths that are two orders of
magnitude larger than the ones expected from lifetime broadening. To highlight this feature, we
have plotted in Fig.~3 a comparison of the theoretical PL line shape for a QD interacting with 1D
and 3D phonon reservoirs. Coupling to 2D or 3D acoustic phonons yields a superohmic spectral
function \cite{Wilson}, for which the independent boson model predicts a (lifetime-broadened) ZPL
accompanied by phonon sidebands (Fig.~3, upper right). This is in strong contrast to our
observations from SWNTs. \footnote[2]{We note, however, that the observed line shapes could also be
reproduced using a subohmic spectral function that would be relevant for confined excitons coupled
to 1D bending modes. The presence of such a coupling on the other hand, requires the breaking of
the circumferential symmetry of the confined excitonic wave function.}

The enhanced pure dephasing of QD transitions in  1D structures was analyzed theoretically in Ref.
\cite{qwire} where the finite width was attributed to the high temperature (Markovian) limit. This
part is due to the temperature-dependent contribution $\chi_{T}(t)$ in Eq.(\ref{eqchiT}) and
vanishes linearly as $T\rightarrow 0$. It provides a symmetric contribution to the line shape (as
illustrated in Fig.~\ref{fignonmarkov}, lower panel, red curve). In contrast, the highly asymmetric
character of the observed line shapes at low temperatures is largely due to the
temperature-independent part of the susceptibility, $\chi_{0}(t)$, in Eq.(\ref{eqchiT0}). For
comparison, we plot these two contributions separately, although they appear in the line shape as a
convolution. The temperature-dependent term is responsible for the presence of a high-energy tail
at non-zero temperature, while the broad low-energy tail we measure constitutes a direct signature
of the temperature-independent pure dephasing. Furthermore, we  find that the characteristic energy
of the latter quasi-exponential tail is directly proportional to the width of the form factor in
\emph{q}-space. This dependence is to be expected since the inverse localization length scale
determines the highest phonon energies to which the QD is effectively coupled. A stronger
confinement therefore leads to a faster dephasing, even at arbitrarily low temperatures.

In Fig.~3 (lower panel) the inset shows the time dependence of the calculated linear susceptibility
$\chi(t)$ used to fit the experimental spectrum: the deviation from an exponential decay that is
apparent for $t \le 1$~psec is a hallmark of non-Markovian decoherence \cite{Loss-Divincenzo}. Our
calculations indicate that this initial (Gaussian) decay crosses over to a power-law decay and
finally to an exponential decay at long times. The regime where power-law decay is dominant expands
as temperature is decreased; further reduction of the temperature to 0.1K leads to the appearance
of power-law decay of the susceptibility extending to 15 psec (blue curve). Observation of such a
strong deviation from Markovian dynamics should be within reach.

To conclude, we have used an extension of the independent boson model to reproduce with remarkable
quantitative accuracy the asymmetric and broad lines that we measured on isolated SWNTs. The model
also reveals the changes in the shape and width with temperature. In strong contrast to 3D systems,
the reduced dimensionality of the phonon bath in SWNTs leads to ohmic coupling of confined excitons
to stretching mode phonons: in this limit, the ZPL is replaced by a power-law singularity. This
conclusion remains valid at arbitrarily low temperatures, where the induced pure dephasing becomes
manifestly non-Markovian, and would apply to QDs defined in any one-dimensional system. These
intrinsic effects are of fundamental interest in view of applications in quantum information
processing, where non-Markovian nature of decoherence may substantially alter the attainable
accuracy thresholds \cite{Burkard}.

The Authors acknowledge enlightening discussions with Ignacio Wilson-Rae. This work was supported
by a grant from the Swiss National Science Foundation (SNSF).

\end{document}